\documentclass[a4paper,12pt]{article}
\newcommand{\bea}{\begin{eqnarray}}
\newcommand{\ena}{\end{eqnarray}}

\begin{document}
\topmargin 0pt \oddsidemargin 0mm

\renewcommand{\thefootnote}{\fnsymbol{footnote}}

\begin{titlepage}

\begin{flushright}
KU-TP 041 \\
\end{flushright}

\vspace{5mm}
\begin{center}
{\Large \bf  Friedmann Equations from Entropic Force}
\vspace{15mm}

{\large Rong-Gen Cai$^{a,}$~\footnote{e-mail: cairg@itp.ac.cn},
Li-Ming Cao$^{b,}$\footnote{e-mail: caolm@phys.kindai.ac.jp}, and
Nobuyoshi Ohta$^{b,}$\footnote{e-mail: ohtan@phys.kindai.ac.jp}}

\vspace{10mm} {\em $^a$  Key Laboratory of Frontiers in
Theoretical
Physics, Institute of Theoretical Physics, \\
Chinese Academy of Sciences, P.O. Box 2735, Beijing 100190, China
\\
 $^b$ Department of Physics, Kinki University, Higashi-Osaka, Osaka
577-8502, Japan}

\end{center}

\vspace{50mm} \centerline{{\bf{Abstract}}} \vspace{5mm}
 In this note by use of the holographic principle together with
 the equipartition law of energy and the Unruh temperature, we
 derive the Friedmann equations of a Friedmann-Robertson-Walker universe.

\end{titlepage}

\newpage
\renewcommand{\thefootnote}{\arabic{footnote}}
\setcounter{footnote}{0} \setcounter{page}{2}
It has a long history that gravity is not regarded as a fundamental
interaction in Nature. The earliest idea on this was proposed by
Sakharov in 1967~\cite{Sak}. In this so-called induced gravity,
spacetime background emerges as a mean field approximation of
underlying microscopic degrees of freedom, similar to hydrodynamics
or continuum elasticity theory from molecular physics~\cite{Visser}.
This idea has been further developed since the discovery of the
thermodynamic properties of black hole in 1970s. Black hole
thermodynamics tells us that a black hole has an entropy
proportional to its horizon area and a temperature proportional to
its surface gravity at the black hole horizon, and the entropy and
temperature together with the mass of the black hole satisfy the
first law of thermodynamics~\cite{Bek,Haw,BCH}.

The geometric feature of thermodynamic quantities of black hole
leads Jacobson to ask an interesting question whether it is possible
to derive Einstein's equations of gravitational field from a
point of view of thermodynamics~\cite{Jacob}. It turns out that it
is indeed possible. Jacobson derived Einstein's equations by
employing the fundamental Clausius relation $\delta Q =TdS$ together
with the equivalence principle. Here the key idea is to demand that
this relation holds for all the local Rindler causal horizon through
each spacetime point, with $\delta Q$ and $T$ interpreted as the
energy flux and Unruh temperature seen by an accelerated observer
just inside the horizon. In this way, Einstein's equation is
nothing but an equation of state of spacetime.

Further, assuming the apparent horizon of a
Friedmann-Robertson-Walker (FRW) universe has temperature $T$ and
entropy $S$ satisfying $T = 1/2\pi\tilde r_A $ and $S = A/ 4G$,
where $\tilde r_A$ is the radius of the apparent horizon and $A$ is
the area of the apparent horizon, one is able to derive Friedmann
equations of the FRW universe with any spatial curvature by applying
the Clausius relation to apparent horizon~\cite{CK}. This works not
only in Einstein's gravitational theory, but also in Gauss-Bonnet and
Lovelock gravity theories. Here a key ingredient is to replace the
entropy area formula in Einstein's theory by using entropy expressions
of black hole horizon in those higher order curvature theories.
Recently the Hawking temperature associated with the apparent
horizon of FRW universe has been shown~\cite{CCH}. There exist a lot
of papers investigating the relation between the first law of
thermodynamics and the Friedmann equations of FRW universe in
various gravity theories. For more references see, for example
\cite{CCH08,Cai07} and references therein.

Another hint appears on the relation between thermodynamics and
gravitational dynamics by investigating the relation between the
first law of thermodynamics and gravitational field equation in the
setup of black hole spacetime. Padmanabhan~\cite{Pad} first noticed
that the gravitational field equation in a static, spherically
symmetric spacetime can be rewritten as a form of the ordinary first
law of thermodynamics at a black hole horizon. This indicates that
Einstein's equation is nothing but a thermodynamic identity. This
observation was then extended to the cases of stationary
axisymmetric horizons and evolving spherically symmetric horizons in
Einstein's gravity~\cite{KSP}, static spherically symmetric
horizons~\cite{PSP} and dynamical apparent horizons~\cite{CCHK} in
Lovelock gravity, and three dimensional BTZ black hole
horizons~\cite{Akbar}. Very recently it has been shown it also holds
in Horava-Lifshitz gravity~\cite{CO}. For a recent review on this
topic and some relevant issues, see \cite{Pad09}.

In a very recent paper by Verlinde~\cite{Verl}, the viewpoint of
gravity being not a fundamental interaction has been further
advocated. Gravity is explained as an entropic force caused by
changes in the information associated with the positions of material
bodies. Among various interesting observations made by Verlinde,
here we mention two of them. One is that with the assumption of the
entropic force together with the Unruh temperature~\cite{Unruh},
Verlinde is able to derive the second law of Newton. The other is
that the assumption of the entropic force together with the
holographic principle and the equipartition law of energy leads to
Newton's law of gravitation. Similar observations are also made by
Padmanabhan~\cite{Pad12}. He observed that the equipartition law of
energy for the horizon degrees of freedom combing with the
thermodynamic relation $S=E/2T$, also leads to Newton's law of
gravity, here $S$ and $T$ are thermodynamic entropy and temperature
associated with the horizon and $E$ is the active gravitational mass
producing the gravitational acceleration in the
spacetime~\cite{Pad03}. Finally we mention that there exist some
earlier attempts to build microscopic models of spacetime, for
example,  see~\cite{in1, in2,in3}.

In this short note we are going to derive the Friedmann equations
governing the dynamical evolution of the FRW universe from the
viewpoint of entropic force together with the equipartition law of
energy and the Unruh temperature by generalizing some arguments of
Verlinde to dynamical spacetimes.

Consider the FRW Universe with metric
\begin{equation}
\label{eq1}
ds^2=- dt^2 +a^2(t)(dr^2 +r^2 d\Omega^2),
\end{equation}
where $a(t)$ is the scale factor of the universe. Following
\cite{Verl}, consider a compact spatial region ${\cal V}$ with
a compact boundary ${\partial \cal V}$, which is a sphere
with physical radius $\tilde r= a r$. The compact boundary
$\partial \cal V$ acts as the holographic screen. The number of
bits on the screen is assumed as
\begin{equation}
\label{eq2}
N = \frac{Ac^3}{G\hbar},
\end{equation}
where $A$ is the area of the screen (note that there is a factor
difference $1/4$ from the Bekenstein-Hawking area entropy formula of
black hole). Assuming the temperature on the screen is $T$, and then
according to the equipartition law of energy, the total energy on
the screen is
\begin{equation}
\label{eq3}
E = \frac{1}{2} N k_B T.
\end{equation}
Further just as in \cite{Verl}, we need the relation
\begin{equation}
\label{eq4}
E=Mc^2,
\end{equation}
where $M$ represents the mass that would emerge in the compact
spatial region ${\cal V}$ enclosed by the boundary screen
$\partial {\cal V}$.

Suppose the matter source in the FRW universe is a perfect fluid
with stress-energy tensor
\begin{equation}
\label{eq5}
T_{\mu\nu}=(\rho +p) u_{\mu} u_{\nu} +pg_{\mu\nu}.
\end{equation}
Due to the pressure, the total mass $M=\rho V$ in the region enclosed by
the boundary $\partial {\cal V}$ is no longer conserved, the
change in the total mass is equal to the work made by the pressure
$dM =-pdV$, which leads to the well-known continuity equation
\begin{equation}
\label{eq6}
\dot \rho +3H (\rho+p)=0,
\end{equation}
where $H=\dot a/a$ is the Hubble parameter.

The total mass in the spatial region  ${\cal V}$ can be expressed as
\begin{equation}
\label{eq7}
M =\int_{\cal V}dV(T_{\mu\nu}u^{\mu}u^{\nu}),
\end{equation}
where $T_{\mu\nu}u^{\mu}u^{\nu}$ is the energy density measured by
a comoving observer. On the other hand, the acceleration for a
radial comoving observer at $r$, namely at the place of the
screen, is
\begin{equation}
\label{eq8}
a_r =-d^2\tilde r/dt^2=-\ddot{a}\ r,
\end{equation}
where the negative sign arises because we consider the acceleration is
caused by the matter in the spatial region enclosed by the boundary
${\partial \cal V}$.
Note that the proper acceleration vanishes for a comoving observer.
However, the acceleration (\ref{eq8}) is crucial in the following
discussions.
According to the Unruh formula, we assume that the acceleration corresponds to a
temperature
\begin{equation}
\label{eq9}
T = \frac{1}{2\pi k_B c}\hbar a_r.
\end{equation}
Now it is straightforward to derive the following equation from
eqs.~(\ref{eq2}), (\ref{eq3}), (\ref{eq4}), (\ref{eq7}) and (\ref{eq9}):
\begin{equation}
\label{eq10}
\ddot a = -\frac{4\pi G}{3} \rho a.
\end{equation}
This is nothing but the dynamical equation for Newtonian cosmology
(Page 10 in \cite{Muk}).  Note that the reference \cite{Muk} derives
(\ref{eq10}) from the Newtonian gravity law, while we obtain
(\ref{eq10}) by using the holographic principle and the
equipartition law of energy in statistical physics.  To produce the
Friedmann equations of FRW universe in general relativity, let us
notice that producing the acceleration is the so-called active
gravitational mass ${\cal M}$~\cite{Pad03}, rather than the total
mass $M$ in the spatial region ${\cal V}$. The active gravitational
mass is also called Tolman-Komar mass, defined as
\begin{equation}
\label{eq11}
{\cal M}= 2\int_{\cal V}dV \left( T_{\mu\nu}-\frac{1}{2}T
g_{\mu\nu}\right) u^{\mu}u^{\nu}.
\end{equation}
Replacing $M$ by ${\cal M}$, we have in this case
\begin{equation}
\label{eq12}
\frac{\ddot a}{a}= -\frac{4\pi G}{3}\left(\rho +3p\right).
\end{equation}
This is just the acceleration equation for the dynamical
evolution of the FRW universe. Multiplying $\dot a a$ on both
sides of eq.~(\ref{eq12}), and using the continuity equation
(\ref{eq6}), we integrate the resulting equation and obtain
\begin{equation}
\label{eq13}
H^2 +\frac{k}{a^2}= \frac{8\pi G}{3}\rho.
\end{equation}
Note that $k$ appears in (\ref{eq13}) as an integration constant,
but it is clear that the constant $k$ has the interpretation as
the spatial curvature of the region ${\cal V}$ in the Einstein
theory of gravity. $k=1$, $0$ and $-1$ correspond to a close,
flat and open FRW universe, respectively.

The above discussion can be extended to any spacetime dimension $d
\ge 4$. In that case, the number of bits on the screen is changed
to~\cite{Verl}
\begin{equation}
N = \frac{1}{2} \frac{d-2}{d-3} \frac{Ac^3}{G\hbar},
\end{equation}
the continuity equation becomes $\dot \rho +(d-1)H (\rho+p)=0$, and
the active mass ${\cal M}$ is defined as
\begin{equation}
\label{eq15}
{\cal M}= \frac{d-2}{d-3}\int_{\cal V}dV \left( T_{\mu\nu}-\frac{1}{d-2}T
g_{\mu\nu}\right) u^{\mu}u^{\nu}.
\end{equation}
The acceleration equation (\ref{eq12}) is changed to
\begin{equation}
\label{eq16}
\frac{\ddot a}{a}= -\frac{8\pi G}{(d-1)(d-2)}\left( (d-3)\rho +(d-1)p \right ).
\end{equation}
Integrating (\ref{eq16}) we then have
\begin{equation}
\label{eq17}
H^2 +\frac{k}{a^2}= \frac{16 \pi G}{(d-1)(d-2)}\rho.
\end{equation}
This is just the Friedmann equation of the FRW universe in $d$
dimensions.

Thus we have derived the Friedmann equations of a FRW universe
starting from the holographic principle and the equipartition law of
energy by using Verlinde's argument that gravity appears as an
entropic force. Before we close this note, however, some remarks are
in order.  First it is claimed that Verlinde's arguments are
applicable to any spacetime, but Verlinde mainly discusses the cases
of static and/or stationary spacetimes. In particular, when he
derives Einstein's equation, a time-like Killing vector is
employed. The time-like Killing vector exists for static or
stationary spacetimes, and it does not for a dynamical spacetime.
Here we have applied his arguments to the FRW universe, a special
dynamical spacetime, and obtained the dynamical equations governing
the evolution of the FRW universe. Second, in deriving Newton's law
of gravity, Verlinde considers a spherical surface with a fixed
radius as the holographic screen, and does not take into account the
evolution of the background spacetime itself. This is right since in
Newton's gravity, the background spacetime is a fixed one. In our
case, the holographic screen is a dynamical one, in some sense, so
it can be viewed as a surface of the spherical symmetric dust
matter~\cite{Muk}. The surface evolves due to the selfgravity. Thus,
an observer (or a test particle) on the screen will feel a force
which leads to an acceleration (\ref{eq8}).  The final comment is
concerned with the assumed relation (\ref{eq9}). According to Unruh,
the acceleration could correspond to a local Unruh temperature on
the screen (\ref{eq9}). Note that the acceleration (\ref{eq8}) is
not a proper acceleration; the proper acceleration vanishes for a
comoving observer in the FRW universe. In fact, $a_r$ is just the
acceleration of geodesic deviation vector~\cite{Pad10}. Let us
recall that Verlinde arrives at the second law of Newton starting
from entropic force together with  the Unruh relation (Eq.~(3.8) in
\cite{Verl}), which relates the temperature on the screen to an
acceleration. Note that the second law of Newton is a nonrelativistic
form, where the acceleration has a form $\ddot x$. The
situation is the same as the case of the discussions in the present
paper. Indeed Eq.~(\ref{eq10}) has a nonrelativistic origin.
It is argued by Verlinde that here the Unruh relation should be read
as a formula for the temperature on the screen that is required to
cause the acceleration, not as usual, as the temperature caused by
an acceleration. Therefore the relation (\ref{eq9}) may be regarded
as a working ansatz here. Thus it is a very interesting issue to see
whether there exists such a relation between the Unruh temperature
and the acceleration. To this aim, some useful references are
already available~\cite{Deser}: It is known that Hawking
temperature in de Sitter space and in some black hole spacetimes can
be viewed as a Unruh temperature for a Rindler observer in
higher-dimensional flat spacetimes in which the de Sitter spacetime
and black holes spacetime can be embedded.

{\bf Note added:} When we are in the final stage of writing the
manuscript,  two papers appear \cite{SG,Pad10} in the preprint
archive, which discuss some relevant issues and have some overlap
with our discussions in this paper.

\section*{Acknowledgments}
RGC thanks his colleagues and students at ITP/CAS and T. Padmanabhan
for various discussions on the entropic force.  RGC is supported
partially by grants from NSFC, China (No. 10821504 and No. 10975168)
and a grant from the Ministry of Science and Technology of China
national basic research Program (973 Program) (No. 2010CB833004).
LMC and NO were supported in part by the Grants-in-Aid for
Scientific Research Fund of the JSPS Nos. 20540283 and
21$\cdot$\,09225, and also by the Japan-U.K. Research Cooperative
Program.


\end{document}